\pdfoutput=1
\documentclass[10pt,a4paper]{article}
\usepackage{jheppub,amsmath,color,graphicx,hyperref,slashed,multirow,lmodern,upquote,titlesec}
\usepackage[utf8]{inputenc}
\usepackage[T1]{fontenc}
\usepackage[onehalfspacing]{setspace}
\DeclareUnicodeCharacter{2212}{-}

\hypersetup{bookmarks=true,unicode=true,pdftoolbar=true,pdfmenubar=true,
	pdffitwindow=false,pdfstartview={FitH},pdftitle={},
	pdfauthor={}, pdfsubject={},
	pdfcreator={},pdfproducer={texstudio}, pdfkeywords={},
	pdfnewwindow=true,colorlinks=true,
	linkcolor=blue,citecolor=magenta,filecolor=magenta,urlcolor=cyan}

\newcommand{\be}{\begin{equation}}
\newcommand{\ee}{\end{equation}}
\def\bsp#1\esp{\begin{split}#1\end{split}}
\def\bpm{\begin{pmatrix}}
\def\epm{\end{pmatrix}}
\title{Cross-fertilising extra gauge boson searches at the LHC}

\author[a]{Jack Y. Araz}
\author[b]{\!\!, Mariana Frank}
\author[c]{\!\!, Benjamin Fuks}
\author[d]{\!\!, Stefano Moretti}
\author[b,d]{\! and \"{O}zer \"{O}zdal}

\affiliation[a]{Institute for Particle Physics Phenomenology Durham University, South Road, Durham, DH1 3LE, UK}
\affiliation[b]{Department of Physics, Concordia University, 7141 Sherbrooke St. West, Montreal, Quebec H4B 1R6, Canada}
\affiliation[c]{Laboratoire de Physique Th\'eorique et Hautes Energies (LPTHE), UMR 7589, Sorbonne Universit\'e et CNRS, 4 place Jussieu, 75252 Paris Cedex 05, France}
\affiliation[d]{School of Physics $\&$ Astronomy, University of Southampton, Highfield, Southampton SO17 1BJ, UK}
\emailAdd{jack.araz@durham.ac.uk}
\emailAdd{mariana.frank@concordia.ca}
\emailAdd{fuks@lpthe.jussieu.fr}
\emailAdd{s.moretti@soton.ac.uk}
\emailAdd{ozer.ozdal@soton.ac.uk}

\abstract{
For the purpose of cross-fertilising currently separate experimental approaches, we connect results of LHC analyses attempting to access the properties of additional $W^\prime$ and $Z^\prime$ bosons from Drell-Yan processes. Under theoretical assumptions linking the two new gauge bosons, we take into account that such possible states of nature are wide enough (in relation to the leptonic mass resolution) for the corresponding signals be significantly affected by interference effects with the background from the Standard Model. The shape of the differential cross section may then no longer be a standard Breit-Wigner distribution, and asymmetry observables would become useful for characterisation (and, possibly, discovery) purposes. Under such conditions we concentrate our analysis on specific widely-studied models: the Sequential Standard Model, a model with an additional $SU(2)_L$ gauge symmetry, as well as standard and alternative realisations of the Left-Right Symmetric Model. We show how information gathered in $Z'$ boson searches in terms of cross section and/or asymmetry distributions can be used to improve $W'$ boson searches in terms of the LHC sensitivity, and {\it vice versa}.
}

\begin{document}
\begin{flushright}
IPPP/21/04
\end{flushright}

\maketitle
\flushbottom

%%%%%%%%%%%%%%%%%%%%%%%%%%%%%%%%%%%%%%%%%%%%%%%%%%%%%%%%%%%%%%%%%%%%%%%%%%%%%%
\section{Introduction}
\label{sec:intro}
The Standard Model (SM) is very successful in explaining most of the available experimental data, and it has survived all tests at colliders. Yet it appears to be theoretically inconsistent or incomplete, and phenomena such as dark matter, neutrino oscillations and matter-antimatter asymmetry remain unanswered. A more complete description of nature, such as via Grand Unification Theories (GUTs), supersymmetry, extra-dimensional models or scenarios featuring dynamical symmetry breaking at a higher scale, is thus in order and should introduce additional symmetries and particles. In particular, extra gauge symmetries predict the presence of extra gauge bosons, with properties similar to those of the usual $Z$ and $W$ bosons of the SM. Specifically, $Z^\prime$ bosons point to the presence of at least an extra $U(1)^\prime$ symmetry \cite{Hewett:1988xc,Langacker:2008yv,Accomando:2016eom}, while $W^\prime$ bosons are indicative of at least an extra $SU(2)^\prime$ gauge group \cite{Mohapatra:1974gc,Senjanovic:1975rk,Beg:1977ti,Ma:1986we}.

Searches for these particles, thus, continue to be an important part of the LHC physics programme. Consequently, an ability to improve the experimental sensitivity in testing theoretical hypotheses is still desirable. Such heavy resonances with significant couplings to quarks and leptons are most easily seen at hadron colliders in the Drell-Yan production mode, wherein quark-antiquark annihilations into a possibly off-shell $Z'$ or $W'$ boson is followed by the extra boson decay into either a dilepton or a lepton plus missing transverse energy final state respectively~\cite{ATLAS:2019erb,ATLAS:2019lsy,Sirunyan:2021khd}. The signals searched for in these Drell-Yan channels are among the cleanest ones at high-energy hadron colliders. Historically, both the charged current and neutral current channels have been pursued separately without exploiting at the experimental level the possibility that $Z'$ and $W'$ states could originate from a single theoretical framework. This would however offer the possibility of optimising searches for one state based on results concerning the other. While such correlations could already be used at the level of total rates to identify the underlying theoretical assumptions (that include at least additional $SU(2)^\prime$ symmetries)~\cite{Jezo:2012rm,deBlas:2012qp}, we push forward this idea and assess the improvements that could arise from potential measurements at the differential level.

The LHC collaborations have already looked for hints of additional gauge bosons, albeit by separately considering neutral and charged states, and imposed stringent constraints on their masses. These states are now required to be heavier than 4--5 TeV, the exact limits depending on the specific new gauge boson properties and those of the model in general (mass $M$, width $\Gamma$, and couplings to the SM states). At such high masses, these resonances have in particular a width that is comparable to the experimental resolution in the invariant and the transverse masses of the final state which are used to extract the new gauge boson signals. This is not only the case when the extra bosons are broad, but also when they are narrow ({\it i.e.}\ for $\Gamma/M\sim \alpha$, with $\alpha$ being the electromagnetic coupling constant) or equivalently when the narrow-width approximation (NWA) holds. Replacing the Breit-Wigner (BW) distribution by a $\delta$ function is therefore an inaccurate way to characterise the production and decay of the new states~\cite{Accomando:2011eu,Accomando:2013sfa,Accomando:2016mvz,Accomando:2019ahs}. Furthermore, in these circumstances,  interference effects with the SM background become visible in the aforementioned mass spectra \cite{Accomando:2013sfa,Kahlhoefer:2019vhz}. A combined correlation of $W^\prime$ and $Z^\prime$ analyses, including finite width effects, is so far lacking. This is addressed in this work.

One way to take into account all correlations between $W^\prime$ and $Z^\prime$ processes in full generality would be to consider a general framework in which the interactions of new heavy resonances are parameterised in a model-independent manner~\cite{Fuks:2017vtl}. In such a scenario, the Lagrangian describing the interactions however includes many free parameters, rendering this task unpractical. Alternatively, it may be desirable to restrict this freedom and choose several simpler and well-motivated specific models. For example, in the Sequential Standard Model (SSM) \cite{Gopalakrishna:2010xm,Han:2012vk}, the $Z^\prime$ and $W^\prime$ couplings are assumed to be SM-like and the only free parameter is their mass, possibly together with a global normalisation factor for the coupling strengths. In models with additional $SU(2)^\prime$ gauge groups~\cite{Hsieh:2010zr}, that include  Left-Right Symmetric Models (LRSMs)~\cite{Pati:1974yy,Mohapatra:1974gc,Senjanovic:1975rk,Mohapatra:1977mj,Senjanovic:1978ev,Ma:2010us}, the  $W^\prime$ and $Z^\prime$ bosons are described in terms of a minimal set of parameters. In this work, we follow this second approach and we investigate the LHC phenomenology of a class of realistic theories Beyond the SM (BSM), with the electroweak symmetry being extended to include an additional $SU(2)'$ gauge group. This predicts the existence of both a $W^\prime$ and a $Z^\prime$ boson, and we will consider the best way to discover those neutral and charged resonances through their Drell-Yan production and decay and by using the invariant mass and  transverse mass distributions of the final-state products. We will consider not only total rates, as done in previous works, but also differential distributions. The signature of the production of a new vector boson and its subsequent decay will indeed lead to a resonant peak in the invariant mass distribution of the dilepton final state (for the $Z'$ boson case) and to a Jacobian shoulder in the transverse mass spectrum of the lepton-neutrino one (for the $W'$ boson case). We will also be looking at the asymmetries of the final states, as the latter observables have already been proven to aid  a potential discovery of wide $W^\prime$ and $Z^\prime$ states~\cite{Accomando:2015cfa,Accomando:2019ahs,Fiaschi:2021okg}.

In the next section we describe the Lagrangian terms relevant to $W^\prime$ and $Z^\prime$ production and decay at the LHC via Drell-Yan processes, and detail specific BSM realisations featuring these extra gauge bosons simultaneously (namely the SSM, a model with an additional $SU(2)_L$, and other models with an additional $SU(2)_R$ symmetry). We also include, for comparison, models featuring extra $U(1)$ gauge groups that are motivated by an $E_6$ GUT symmetry or by gauging $B-L$ number conservation. We then discuss experimental limits on $W^\prime$ and $Z^\prime$ bosons and next present our numerical results in Section~\ref{Results}. We summarise our work and conclude in Section~\ref{sec:conclusion}.

%%%%%%%%%%%%%%%%%%%%%%%%%%%%%%%%%%%%%%%%%%%%%%%%%%%%%%%%%%%%%%%%%
\section{Theoretical frameworks}\label{sec:theo}
In this section, we briefly introduce the various theoretical frameworks relevant to our study. We begin with a general Lagrangian description, before moving on with details on the various well-motivated models considered. In the last part of this section, we introduce $U(1)'$ extensions of the SM which we will use as reference standards for the comparison of various LHC signals.

\subsection{Models with additional $W'$ and $Z'$ bosons}
\subsubsection{General Lagrangian description}\label{sec:generallag}
In this study, we examine the LHC phenomenology of SM extensions featuring an enlarged gauge group. We assume that the SM gauge symmetry is supplemented by an additional generic $SU(2)^\prime$ group, which predicts the existence of both $W^\prime$ and $Z^\prime$ bosons. By considering the couplings to the fermions to be arbitrary, their form is only dictated by Lorentz and gauge invariance. The most general  charged current (CC) and neutral current (NC) Lagrangians ${\cal L}_{\rm CC}^{W^\prime}$ and ${\cal L}_{\rm NC}^{Z^\prime}$ respectively describing the interactions of a charged and a neutral spin-one gauge boson with the SM fermions $f$ then read, after including right-handed neutrino fields,
\be\bsp
  {\cal L}_{\rm CC}^{W^\prime}=&\ \frac{g_2^\prime}{\sqrt{2}} \left[ {\bar u}_i \gamma^\mu \left( [\kappa_{q,L}^{W^\prime}]_{ij}P_L +[\kappa_{q,R}^{W^\prime}]_{ij}P_R\right) d_j  + {\bar \nu}_i \gamma^\mu \left( [\kappa_{\ell,L}^{W^\prime}]_{ij}P_L +[\kappa_{\ell,R}^{W^\prime}]_{ij}P_R\right) \ell_j \right] W^\prime_\mu + {\rm H.c.}\, , \\
{\cal L}_{\rm NC}^{Z^\prime}=&\ \frac{g_2^\prime}{\cos \theta_W} \left [{\bar q}_i \gamma^\mu \left( [\kappa_{q,L}^{Z^\prime}]_{ij}P_L +[\kappa_{q,R}^{Z^\prime}]_{ij}P_R\right) q_j  + {\bar \ell}_i \gamma^\mu \left( [\kappa_{\ell,L}^{Z^\prime}]_{ij}P_L +[\kappa_{\ell,R}^{Z^\prime}]_{ij}P_R\right) \ell_j \right] Z^\prime_\mu + {\rm H.c.} \, ,
\esp\label{eq:LagrangianMI}\ee
where a summation over quark and lepton flavours is understood for both Lagrangians, and where we denote by $g_2^\prime$ the $SU(2)^\prime$ coupling constant. Here, the arbitrary parameters $\kappa$ are complex-valued elements of $3\times3$ matrices in the flavour space (with $q=u, d$ and $i,j$ being generation indices). These expressions for the CC and NC Lagrangians are inspired by the SM, that we recover by setting
\be\bsp
  [\kappa_{q,L}^{W^\prime}]_{ij}    = V_{ij}^{\rm CKM} \, , \qquad
  [\kappa_{\ell,L}^{W^\prime}]_{ij} = V_{ij}^{\rm PMNS}\, , \qquad
  \kappa_{q (\ell),R}^{W^\prime} = 0 \, ,\\
  [\kappa_{f,L}^{Z^\prime}]_{ij} = \left(T_{L}^{3,f}-Q_f\sin^2\theta_W \right) \delta_{ij} \, ,\qquad
  [\kappa_{f,R}^{Z^\prime}]_{ij} = \left(-Q_f\sin^2\theta_W \right) \delta_{ij}\, ,
\esp\label{eq:SMassignments}\ee
after assuming that the $g_2'$ coupling is the same as the weak coupling constant $g_W$. In this notation, $V^{\rm CKM}$ and $V^{\rm PMNS}$ are the CKM and PMNS matrices, $T^3_L$ and $Q$ are the weak isospin and electric charge operators and $\theta_W$ is the electroweak mixing angle.  Whereas in principle, gauge boson mixing can induce $W'$ and $Z'$ couplings to a pair of SM weak bosons, such mixings are either suppressed or forbidden in the new physics frameworks considered. They are therefore ignored in the following.

These couplings encapsulate a large number of free parameters, and this still holds even if we rely on sum rules, unitarity constraints and anomaly cancellations to reduce the number of degrees of freedoms. As intimated,  a completely model-independent study of the associated collider phenomenology is thus unfeasible practically, although attempts in this direction have been made~\cite{deBlas:2012qp}. Furthermore, a detailed analysis of a particular chosen model configuration, where we fix {\it some} of the parameters arbitrarily, may not correspond to a theoretically-motivated framework and thus not be very informative. For this  reason, we rely in the following on specific classes of models predicting $W^\prime$ and $Z^\prime$ bosons couplings parameterised by a limited set of common free parameters. Our findings and our method, that we will describe below, can however easily be generalised to other BSM theories.

%%%%%%%%%%%%%%%%%%%%%%%%%%%%%%%%%%%%%%%%%%%%%%%%%%%%%%%%%%%%%%%%%%%%%%%%%%%%%%

\subsubsection{The Sequential Standard Model}\label{sec:modelSSM}
The simplest approach to model the dynamics associated with additional gauge bosons would be to extend the SM field content by two massive $W^{\prime}$ and $Z^\prime$ vector fields that are charged and neutral respectively, as in Section~\ref{sec:generallag}. The form of the $W^\prime$ and $Z^\prime$ chiral couplings to all SM fermions is then imposed to be the same as in the SM, and their possible interactions with other gauge and scalar bosons are omitted. This approach, which is traditionally coined the SSM~\cite{Gopalakrishna:2010xm,Han:2012vk}, is described by the model Lagrangian~\eqref{eq:LagrangianMI} with the couplings~\eqref{eq:SMassignments}. Moreover, as no right-handed neutrinos are present in the SM, the corresponding right-handed leptonic couplings are all zero. The $W^\prime$ and $Z^\prime$ bosons being introduced in an {\it ad-hoc} manner, their masses (and sometimes their widths even if they are in principle calculable) are thus the only new physics free parameters. In this work, we additionally take the new physics interactions to be flavour-diagonal for simplicity.
%%%%%%%%%%%%%%%%%%%%%%%%%%%%%%%%%%%%%%%%%%%%%%%%%%%%%%%%%%%%%%%%%%%%%%%%%%%%%%

\subsubsection{Models with an additional $SU(2)_L$}\label{sec:modelLL}
A more consistent way to add simultaneously extra gauge bosons to the SM field content is to enlarge the electroweak symmetry group. For this purpose, we consider the group structure $SU(2)_1 \times SU(2)_2 \times U(1)_X$ related to the commonly called $G(221)$ models~\cite{Hsieh:2010zr}. Models of this class all feature additional $W^\prime$ and $Z^\prime$ bosons, the various possibilities differing through the quantum number assignments of the different fields. We examine a configuration in which the $SU(2)_L$ symmetry of the SM is doubled, so that $SU(2)_1 \equiv SU(2)_L$ and $SU(2)_2 \equiv SU(2)_L'$. The representation of the weak doublets of SM fermions then read
\be
 L_{L} = \begin{pmatrix}\nu_L \\ \ell_L\end{pmatrix} \, \sim (\mathbf{2},\mathbf{2},X_Q) \, ,\qquad
 Q_{L} = \begin{pmatrix}u_L \\ d_L\end{pmatrix} \, \sim (\mathbf{2},\mathbf{2},X_L) \, ,
\ee
where the $U(1)_X$ quantum numbers $X_Q$ and $X_L$ are fixed to recover the correct hypercharge quantum numbers once the extended gauge symmetry is broken down to the electroweak one, $SU(2)_L \times SU(2)_L^\prime \times U(1)_X \to SU(2)_L \times U(1)_Y$. Here, the right-handed fermions are all singlets under the two $SU(2)$. In a second step, the electroweak symmetry is further broken to the electromagnetic gauge symmetry $U(1)_{\rm EM}$. Such a two-stage breaking is achieved through a Higgs sector containing a scalar bidoublet $\Phi$, that implies $SU(2)_L^\prime \times U(1)_X \to U(1)_Y$, and the usual SM weak doublet $H$, that leads to $SU(2)_L \times U(1)_Y\to U(1)_{\rm EM}$. Those scalar fields are given, in terms of their component fields, by
\begin{equation}
 \Phi \equiv \frac {1}{\sqrt{2}}\begin{pmatrix} \phi^0  & \sqrt{2} \phi^+\\ \sqrt{2} \phi^-& \phi^0 \end{pmatrix} \sim (\mathbf{2},\mathbf{2^*},\mathbf{1/2})\, , \qquad
 H \equiv \begin{pmatrix} h^+  \\  h^0 \end{pmatrix} \sim (\mathbf{1},\mathbf{2},\mathbf{1/2})\, ,
\end{equation}
and the symmetry is broken once the neutral components of these fields get non-vanishing vacuum expectation values (vevs), $\langle \phi^0\rangle = u$ and $\langle h^0\rangle = v/\sqrt{2}$ with $u \gg v$ for consistency with experimental data.

In such a realisation, the physical gauge bosons are admixtures of the $SU(2)_L'$, $SU(2)_L$ and $U(1)_X$ bosons, and their masses are not independent parameters. These can be derived from the vevs of the Higgs fields and the gauge coupling constants $g_1\equiv g_W$, $g_2\equiv g_2'$ and $g_X$,
 \be\bsp
M^2_{W^\prime} =&\ \frac14 (g_1^2+g_2^2) u^2+\frac{\sin^2 \phi}{4} g_1^2 v^2 +\sin^4 \phi  \frac{g^2_1g^2_2}{4(g_1^2+g_2^2)}\frac{v^4}{u^2}\\
M^2_{Z^\prime} =&\ \frac14 (g_1^2+g_2^2) u^2+\frac{\sin^2 \phi}{4} g_1^2 v^2 -\sin^4 \phi \frac{g^2_X}{4e^2} \frac{g^2_1g^2_2}{(g_1^2+g_2^2)}\frac{v^4}{u^2}
\esp \qquad\text{with}\qquad\tan\phi=\frac{g_X}{g_2}\, .
\label{eq:ZpWpmassLL}\ee
The extra gauge bosons are thus practically degenerate. The mixing relations among them also fix their couplings with the SM fermions. They are written, in the notation of Eq.~\eqref{eq:LagrangianMI}, as
\be\bsp
    \kappa_{q, L}^{W^\prime} = \frac{1}{\tan \theta_W}-\cos\theta_W\ \zeta\Big(\frac{\tan\theta_W}{M^2}\Big),
    \qquad
 &  \kappa_{q, R}^{W^\prime} =0,\\
    \kappa_{\ell, L}^{W^\prime}=- \tan \theta_W-\cos\theta_W \ \zeta\Big(\frac{\tan\theta_W}{M^2}\Big), \qquad
 &  \kappa_{\ell, R}^{W^\prime} =0,\\
    \kappa_{q, L}^{Z^\prime} =  T_{3f}\frac{\cos \theta_W}{\tan\theta_W}+\left (T_{3f} +Q_f \sin^2 \theta_W \right) \ \zeta\Big(\frac{\tan\theta_W}{M^2}\Big), \qquad
 &  \kappa_{q, R}^{Z^\prime} =Q_f \sin^2 \theta_W \ \zeta\Big(\frac{\tan\theta_W}{M^2}\Big) ,\\
    \kappa_{\ell, L}^{Z^\prime} = -T_{3f} \sin \theta_W + \left (T_{3f} +Q_f \sin^2 \theta_W \right) \ \zeta\Big(\frac{\tan\theta_W}{M^2}\Big) , \qquad
 &  \kappa_{\ell, R}^{Z^\prime} =Q_f \sin^2 \theta_W   \ \zeta\Big(\frac{\tan\theta_W}{M^2}\Big).
\esp\ee
These couplings depend on the sum of the two weak isospin quantum numbers, $T_{3f}=T_{L}^{ 3,f}+ T_{L}^{\prime \, 3,f}$, as well as on the electroweak mixing angle defined by
\be
 \tan \theta_W=\frac{g_Y}{g_1}\qquad\text{with}\qquad \frac{1}{g_Y^2}=\frac{1}{g_2^2}+\frac{1}{g_X^2}.
\ee
Moreover, $M$ stands for the approximately degenerate mass of the $W^\prime$ and $Z^\prime$ bosons and the function $\zeta (\tan\theta_W/M^2)$ involves subleading corrections of ${\cal O}(\tan \theta_W/M^2)$. In this class of models, the $SU(2)$ bosons only couple to the left-handed fermions. The right-handed couplings are thus either zero (for the $W'$ boson) or strongly suppressed by mixing (for the $Z'$ boson).

Electroweak precision tests constrain the masses of these bosons to be greater than 2.5~TeV~\cite{Hsieh:2010zr}, and we will address limits originating from direct searches at colliders below.

%%%%%%%%%%%%%%%%%%%%%%%%%%%%%%%%%%%%%%%%%%%%%%%%
\subsubsection{Left-Right Symmetric Models}\label{subsec:LRSM}
In this section we still consider the $G(221)$ framework, but this time we enforce the $SU(2)_2$ group to be an $SU(2)_R$ gauge symmetry, and the $U(1)_X$ symmetry to be associated with $B-L$ conservation (with $B$ being the baryon and $L$ the lepton number). The enlarged gauge group is thus $ SU(2)_L\times SU(2)_R \times U(1)_{B-L}$. This setup gives rise to the commonly called LRSM~\cite{Pati:1974yy,Mohapatra:1974gc,Senjanovic:1975rk,Mohapatra:1977mj,Senjanovic:1978ev} in which the left-handed fermions of the SM are organised into doublets of $SU(2)_L$ and are singlet under $SU(2)_R$. In contrast, the right-handed fermions (including right-handed neutrinos) are organised into doublets of $SU(2)_R$ and are singlet under $SU(2)_L$. The $B-L$ quantum numbers are then the usual ones, namely 1/6 for quark and $-1/2$ for lepton fields.

Similarly to the previously-discussed $SU(2)_L\times SU(2)_L'\times U(1)_X$ models, the extended gauge symmetry is broken into two stages. The Higgs sector includes first two Higgs triplets $\Delta_L$ and $\Delta_R$ respectively lying in the adjoint representation of $SU(2)_L$ and $SU(2)_R$ and allowing for the preservation of the left-right symmetry. In addition, it contains a Higgs bidoublet $\Phi$ state. The former Higgs fields induce $SU(2)_R \times U(1)_{B-L}\to U(1)_Y$ breaking, whereas the latter yields electroweak symmetry breaking. The Higgs field representation under the extended gauge group and the electric charge assignments of the various component fields are given by
\begin{equation}\bsp
  \Delta_{L} \equiv \begin{pmatrix} \delta_{L}^+/\sqrt{2} & \delta_{L}^{++} \\ \delta_{L}^0 & -\delta_{L}^+/\sqrt{2} \end{pmatrix} \sim (\mathbf{3},\mathbf{1},2) \,   ,\qquad
  \Delta_{R} \equiv \begin{pmatrix} \delta_{R}^+/\sqrt{2} & \delta_{R}^{++} \\ \delta_{R}^0 & -\delta_{R}^+/\sqrt{2} \end{pmatrix} \sim (\mathbf{1},\mathbf{3},2)\, ,\\
  \Phi \equiv \begin{pmatrix} \phi_1^0 & \phi_2^+ \\ \phi_1^- & \phi_2^0 \end{pmatrix} \sim (\mathbf{2},\mathbf{2},0)\, . \hspace*{4cm}
\esp\end{equation}

As in Section~\ref{sec:modelLL}, the masses of the extra gauge bosons and their couplings to the fermions are dictated by the gauge couplings and the vevs of the neutral Higgs fields, $\langle \delta_{L}^0\rangle = v_{L}/\sqrt{2}$, $\langle \delta_R^0\rangle = v_{R}/\sqrt{2}$, $\langle\phi_1^0\rangle = v_1/\sqrt{2}$ and $\langle\phi_2^0\rangle = v_2/\sqrt{2}$ (when ignoring potentially non-zero phases that cannot be re-absorbed into field redefinitions). Experimental data from the kaon and the $B$-meson sector enforce that $v_R \gg v_1, v_2\gg v_L$ and $\sqrt{v_1^2+v_2^2} = v = 246$~GeV, and strongly limit the mixing possibilities between the SM and the extra gauge bosons. Under those assumptions, we obtain for the squared $W'$  and $Z'$ boson masses
\be\bsp
 M_{W^\prime}^2=&\ \frac{1}{4}\bigg[2g_2^2v_R^2+g_2^2v^2+2g_2g_1\frac{v^2_1v^2_2}{v_R^2} \bigg]\, ,\\
 M^2_{Z^\prime} =&\ \frac14 \bigg[ g_1^2 v^2+2v_R^2 (g_2^2+g_X^2) +\sqrt{\Big[g_1^2 v^2+2v_R^2 (g_2^2+g_X^2)\Big]^2-4g_1^2(g_2^2+2g_X^2)v^2v_R^2} \bigg]\, ,
\esp\ee
with $g_1\equiv g_W$, $g_2\equiv g_R$, and $g_X\equiv g_{B-L}$. Expressions for the couplings of the extra gauge bosons to the SM fermions (including right-handed neutrinos) feature contributions from the three gauge groups that are modulated by the various mixing angles related to the gauge sector. They can be written in the form of Eq.~\eqref{eq:LagrangianMI}, provided that all gauge coupling constants are absorbed in the $\kappa$ parameters. We refer to the works~\cite{Roitgrund:2014zka,Mattelaer:2016ynf} for their explicit derivation.

%%%%%%%%%%%%%%%%%%%%%%%%%%%%%%%%%%%%%%%%%%%%%%%%%%%%%%%%%%%%%%%%%%%%%%%%%%%%%

\subsubsection{Alternative Left-Right Model}\label{subsec:ALRSM}
The Alternative Left-Right Symmetric Model (ALRSM) relies on the same gauge group as in Section~\ref{subsec:LRSM}, but the fields are organised differently into multiplets of the gauge group. Such an option emerges from the breaking of an $E_6$ GUT group into its $SU(3)\times SU(3)\times SU(3)$ maximal subgroup. One of these $SU(3)$'s remains unbroken and is associated with the SM strong interaction group, while the two others $SU(3)$ groups further break into $SU(2)_1 \times SU(2)_2 \times U(1)_X$. There are several options for this breaking to occur, and they respectively lead to the Inert Doublet Model~\cite{Deshpande:1977rw,Ma:2006km,Barbieri:2006dq}, the usual LRSM of the previous section, or the ALRSM~\cite{Ma:1986we} considered in this section.

In the ALRSM, the $SU(2)_2\equiv SU(2)^\prime_{R}$ partner of the right-handed up-quark is an exotic down-type quark and the $SU(2)^\prime_{R}$ partner of the right-handed charged lepton is a new neutral lepton dubbed a scotino. The more common right-handed neutrino and down-type quark thus remain singlets under both the $SU(2)_1\equiv SU(2)_L$ and $SU(2)^\prime_{R}$ groups. Furthermore, the model includes $SU(2)_L$ singlet counterparts to the new states, {\it i.e.}\ $SU(2)_L$ scotinos and exotic down-type quarks. Whereas the field including the SM states have standard $B-L$ quantum numbers, the exotic singlets are assigned $B-L$ quantum numbers of 0 and $-2/3$ for the scotino and the down-type quarks respectively.

The breaking of the extended gauge symmetry is again performed in two steps. It relies, this time, on a pair of Higgs doublets $\chi_L$ and $\chi_R$ that break $SU(2)^\prime_{R}\times U(1)_{B-L}$ into the hypercharge group, and on a Higgs bidoublet $\Phi$ that ensures electroweak symmetry breaking. The representation and charge assignments of the various component fields are given by
\begin{equation}\bsp
  \chi_{L}\equiv\begin{pmatrix} \chi_{L}^+\\ \chi_{L}^0\end{pmatrix} \sim (\mathbf{2},\mathbf{1},1)\, ,\qquad
  \chi_{R}\equiv\begin{pmatrix} \chi_{R}^+\\ \chi_{R}^0\end{pmatrix} \sim (\mathbf{1},\mathbf{2},1)\, ,\qquad
  \Phi \equiv \begin{pmatrix} \phi_1^0 & \phi_2^+ \\ \phi_1^- & \phi_2^0 \end{pmatrix} \sim (\mathbf{2},\mathbf{2},0)\, .
\esp\end{equation}
Once again, the masses and mixing of the extra gauge bosons are not free parameters and depend on the vevs of the neutral components of the Higgs states, $\langle \chi_L^0\rangle = v_L/\sqrt{2}$, $\langle \chi_R^0\rangle = v_R/\sqrt{2}$ and $\langle \phi^0_2\rangle = k/\sqrt{2}$. In the ALRSM, the conservation of a generalised lepton number protects $\phi^0_1$ from acquiring any vev, and forbids the mixing of the SM quarks with the exotic quarks, as well as the one that could occur between the charged gauge bosons. The extra gauge boson masses read
\be\bsp
  M_{W'} = &\ \frac12 g_2 \sqrt{k^2+v_R^2}\ ,\\
  M_{Z'} = &\ \frac12 \sqrt{g_{X}^2 v_L^2 {\sin^2 \phi_W}  + \frac{g_2^2 k^2 (\cos^4 {\phi_W}  + v_R^2)}{\cos^2 {\phi_W}}}\ ,
\esp\ee
where as in Section~\ref{subsec:LRSM}, $g_X\equiv g_{B-L}$ and $g_2\equiv g_R$. The mixing angle $\phi_W$ that appear in the expression for the $Z'$ boson mass describes the mixing of the $U(1)_{B-L}$ gauge boson with the neutral $SU(2)_R'$ gauge boson. Such a mixing gives rise to the hypercharge gauge boson and is defined by
\be
  \sin {\phi_W} = \frac{g_X}{\sqrt{g_X^2+g_2^2}} \equiv \frac{g_Y}{g_2}\, .
\ee
It is hence connected to the hypercharge coupling constant $g_Y$.

As for the LRSM, the coupling of the new gauge bosons to fermions can be written in the form of Eq.~\eqref{eq:LagrangianMI} once we absorb the full gauge coupling dependence into the $\kappa$ parameters. We refer to the work of Refs.~\cite{Ashry:2013loa,Frank:2019nid} for details.

%%%%%%%%%%%%%%%%%%%%%%%%%%%%%%%%%%%%%%%%%%%%%%%%%%%%%%%
\subsection{Models with an additional $Z'$ boson}
\label{sec:extrau1}
%%%%%%%%%%%%%%%%%%%%%%%%%%%%%%%%%%%%%%%%%%%%%%%%%%%%%%%%
In our study we compare predictions for models containing both a $W'$ and a $Z'$ boson also with those associated with models solely featuring an extra gauge boson $Z'$. In this case, the absence of any anomaly, in either cross section or charge asymmetry in the $\ell\nu$ final state which would correlate to signal evidences in the $\ell^+\ell^-$ one, could be indicative of the existence of such a $U(1)'$ extra $Z’$ boson. In this section, we briefly review two of the most popular classes of models that feature an extra Abelian gauge symmetry (and that are thus based on the extended $SU(2)_{L}\times U(1)_Y \times U(1)^\prime$ gauge symmetry), namely models that emerge from the breaking of a GUT group (such as $E_6$) and $U(1)_{B-L}$ models.

%%%%%%%%%%%%%%%%%%%%%%%%%%%%%%%%%%%%%%%%%%%%%%%%%%
\subsubsection{$E_6$ motivated $U(1)^\prime$ models}\label{subsec:u1E6models}
%%%%%%%%%%%%%%%%%%%%%%%%%%%%%%%%%%%%%%%%%%%%%%%%%%%%%
\begin{table}
  \begin{center}
  \renewcommand{\arraystretch}{1.6}
  \setlength\tabcolsep{2pt}
  \begin{tabular}{l||c|c|c|c|c|c}
  & $2\sqrt{10}Q^{'}_\chi $ & $ 2 \sqrt{6}Q^{'}_\psi $  & $2\sqrt{15}Q^{'}_\eta$
    & $ 2\sqrt{15}Q^{'}_S $ & $ 2Q^{'}_I $ & $ 2\sqrt{10}Q^{'}_N $ \\
  \hline \hline
  $\theta_{E_6} $ & $ -\pi/2$ & $ 0 $ & $ \arccos\sqrt {\frac{5}{8}} -\pi $ &
    $\arctan\sqrt{\frac{15}{9}} -\frac{\pi}{2}$ & $\arccos\sqrt{\frac{5}{8}} + \frac{\pi}{2} $ &
    $\arctan\sqrt{15} -\frac{\pi}{2}$ \\
  \hline
  $ Q_{Q_L,u_R,e_R} $ & $-1$     & 1  & $-2$ & $-1/2$ & 0  & 1 \\
  $ Q_{d_R,L_L} $ &  3     & 1  & 1  &   4  & $-1$ & 2 \\
  $ Q_N $ & $-5$   & 1  & $-5$ &  $-5$  & 1  & 0 \\ \hline
  $ Q_{H_u} $ & 2  & $-2$ & 4  &  1   & 0  & 2 \\
  $ Q_{H_d} $ & $-2$ & $-2$ & 1  & $-7/2$ & 1  & $-3$ \\
  $ Q_S $ & 0      & 4  & $-5$ &  5/2 & $-1$ & 5 \\
  \end{tabular}
  \caption{$U(1)^\prime$ charges of the quarks, leptons, and Higgs states in six commonly-studied $U(1)'$ extensions of the SM that could arise from the breaking of a $E_6$ GUT symmetry. We additionally present the value of the mixing angle $\theta_{E_6} \in [-\pi,\pi]$.\label{tab:u1charges}}
  \end{center}
\end{table}

GUT models based on the breaking of the exceptional group $E_6$~\cite{Gursey:1975ki,Achiman:1978vg} have been well motivated from developments in string theories~\cite{Hewett:1988xc} and as generators of models featuring additional $U(1)$ symmetries. The so-called $U(1)'$ models can be seen as arising from the breaking of $E_6$ into $SO(10)\times U(1)$, rather than directly into the $SU(3)\times SU(3)\times SU(3)$ subgroup as in Sections~\ref{subsec:LRSM} and \ref{subsec:ALRSM}. The most commonly-used parameterisation to define the extra $U(1)'$ symmetry emerges from a linear combination of the $U(1)^\prime_\psi$ and $U(1)^\prime_\chi$ groups resulting from the breaking $E_6 \to SO(10) \times U(1)^\prime_\psi \to (SU(5) \times U(1)^\prime_\chi) \times U(1)^\prime_\psi$. In this context, the corresponding charge operator $Q^\prime$ is written as an admixture of the initial $U(1)^\prime_\psi$ and $U(1)^\prime_\chi$ charge operators $Q^\prime_\psi$ and $Q^\prime_\chi$,
\be\label{eq:u1charge}
  Q^\prime = Q^\prime_\psi\cos\theta_{E_6} - Q^\prime_\chi\sin\theta_{E_6}\, ,
\ee
in which we introduce the mixing angle $\theta_{E_6}$. Several benchmark models with various mixing angle and quantum numbers for the SM fields have been considered, and are presented in Table~\ref{tab:u1charges}. Variants exist and generally differ by their (possibly secluded and extended) Higgs sector.

We adopt a configuration in which the Higgs sector contains two Higgs doublets $H_u$ and $H_d$, as well as a Higgs singlet $S$ (that is charged under the extra $U(1)'$ symmetry). This choice is a minimal option that is consistent with the supersymmetric version of the model. The representation of those fields are given by
\be
  H_{u}= \left(\begin{array}{c}  H_u^+ \\ H_u^0   \end{array} \right) \sim (\mathbf{2}, 1/2, Q_{H_u})\, ,\qquad
  H_{d}= \left(\begin{array}{c}  H_d^0 \\ H_d^-   \end{array} \right) \sim (\mathbf{2},-1/2, Q_{H_d})\, ,\qquad
  S \sim (\mathbf{1},0,Q_S)\ ,\
\ee
their $U(1)'$ quantum numbers being provided in Table~\ref{tab:u1charges} for the specific $U(1)'$ models considered. The model additionally includes right-handed neutrinos.

After the spontaneous breaking of the symmetry group down to electromagnetism, the $W$, $Z$ and $Z^\prime$ bosons get massive while the photon stays massless. In general, the SM $Z$ and the extra $Z^\prime$ eigenstates mix, although such a mixing is strongly constrained by electroweak precision data. In our study, we consider an important limit in which the extra gauge boson is much heavier than the SM weak bosons and does not mix. Its mass can then be simply related to the weak coupling constant $g_W$, the quantum numbers of the Higgs states, and their vevs $\langle H_u^0 \rangle = v_u/\sqrt{2}$, $\langle H_d^0 \rangle = v_d/\sqrt{2}$ and $\langle S \rangle = v_s/\sqrt{2}$. It is written as
\be
  M_{Z'}^2 = g_W^2 \Big( Q_{H_u} v_u^2 + Q_{H_d} v_d^2 +Q_S^2 v_s^2 \Big)\, ,
\ee
and generalisations from this limiting case are outlined in Ref.~\cite{Langacker:1984dp}.

%%%%%%%%%%%%%%%%%%%%%%%%%%%%%%%%%%%%%%%%%%%%%%%%%%%%%%%
\subsubsection{Models with an extra $U(1)_{B-L}$ }
\label{subsec:uBLmodels}
%%%%%%%%%%%%%%%%%%%%%%%%%%%%%%%%%%%%%%%%%%%%%%%%%%%%%%
As a second class of models featuring an additional neutral gauge boson relative to the SM, we consider frameworks in which the SM gauge symmetry is extended by gauging the $B-L$ number. The gauge symmetry of the model is thus based on the $SU(2)_L \times U(1)_Y \times U(1)_{B-L}$ group~\cite{Basso:2008iv,Basso:2009hf}, and its dynamics is similar to the other models treated in Section~\ref{subsec:u1E6models}. Differences lie at the level of the $U(1)$ charges of the various fields and concern the details of the Higgs sector.

In terms of its field content, the model includes additional right-handed neutrinos relatively to the SM, so that gauge and gravity anomalies cancel. Moreover, it includes an extra Higgs singlet $S$ that allows for the spontaneous breaking of the $B-L$ symmetry. The quantum numbers of all scalar fields are given by
\be
  H= \left(\begin{array}{c}  h^+ \\ h^0   \end{array} \right) \sim (\mathbf{2}, 1/2, 0)\, ,\qquad
  S \sim (\mathbf{1},0,2)\ .\
\ee
In addition to the SM fermion masses, symmetry breaking also yields the generation of neutrino masses through the implementation of a seesaw mechanism. As usual, gauge boson masses can be written in terms of the gauge couplings and the vevs of the two Higgs bosons, $\langle h^0\rangle = v/\sqrt{2}$ and $\langle S \rangle =  v_s/\sqrt{2}$. We obtain, for the heaviest $Z'$ state,
\be
  M_{Z^\prime} = 2 g_{B-L} v_s \ ,
\ee
where we ignored $Z-Z'$ mixing for simplicity.

%%%%%%%%%%%%%%%%%%%%%%%%
\subsection{Constraints from direct searches for additional gauge bosons}\label{sec:WpZpmass}
All extra gauge bosons studied in this work couple to a pair of SM quarks, with the exception of the $W'$ boson in ALRSM scenarios that features mixed couplings to one SM quark and one exotic quark. Limits can therefore be extracted from ATLAS and CMS dijet searches for new resonances in most cases considered. The analysis of the entire ATLAS~\cite{ATLAS:2019fgd} and CMS~\cite{CMS:2019gwf} Run~2 dataset yields respective bounds of 3.6--4~TeV on SSM $W'$ bosons, and of 2.7--2.9~TeV on the corresponding $Z'$ states. These $W'$ limits can readily be extended to models featuring either an extra $SU(2)_L^\prime$ symmetry or an extra $SU(2)_R$ symmetry, at least in situations where all $SU(2)$ gauge couplings are equal ($g_1=g_2\equiv g_W$).

As dijet bounds on extra $Z'$ bosons are quite weak, more stringent bounds could originate from other channels, like the cleaner dilepton probes. The shape analysis of the high-mass dilepton spectrum indeed allows, by making use of the entire LHC Run~2 dataset, for the exclusion of SSM $Z^\prime$ masses ranging up to 5.1~TeV~\cite{Sirunyan:2021khd,ATLAS:2019erb}. Those limits are only slightly reduced in the framework of $E_6$-inspired $U(1)^\prime$ models, with the $Z'$ mass being enforced to be larger than about 4.56~TeV in $U(1)'_\psi$ models for instance. Leptophobic $Z'$ bosons, that we do not consider in this study, are therefore much less constrained~\cite{Araz:2017wbp}. Here, limits originating from resonance searches through top-antitop pair production become more relevant. The analysis of 36.1~fb$^{-1}$ of LHC data constraints (narrow) $Z'$ boson masses to be greater than 3.1--3.8 TeV, the exact bound depending on the details of the model~\cite{Aaboud:2019roo,CMS:2018rkg}.

Bounds have also been extracted from searches for $Z^\prime$ bosons in the ditau channel. They are, however, weaker. The analysis of a luminosity of $36~{\rm fb}^{-1}$ of LHC Run~2 data hence leads to a mass limit of about 2.4~TeV~\cite{Aaboud:2017sjh}.

Finally, several studies are dedicated to LRSM scenarios. They consider two specific signatures of a resonantly-produced $W_R$ boson, a first one when it decays into a $tb$ system~\cite{Sirunyan:2017vkm,ATLAS:2018wmg}, and a second one when it decays into a lepton and a right-handed neutrino~\cite{ATLAS:2018dcj,ATLAS:2019isd,CMS:2018agk}, the latter new physics state further decaying into a lepton plus dijet system. In the former channel, limits on the mass of the $W_R$ boson have been found to be of 3.2--3.6~TeV after relying on 36~fb$^{-1}$ of LHC data. In the latter channel, the analysis of about 36~fb$^{-1}$ of Run~2 LHC data (and 80~fb$^{-1}$ in the boosted regime) reveals constraints on $W_R$ boson masses lying in the 3.5--4.4~TeV range for right-handed neutrino masses below about 3~TeV.

%%%%%%%%%%%%%%%%%%%%%%%%
\section{Phenomenological investigations}\label{Results}
In this section, we investigate various signals of additional (heavy) gauge bosons at the LHC, as well as the correlations that could emerge from the simultaneous presence of additional $W^\prime$ and $Z^\prime$ states in the model. Our aim is to get indications on the potential non-existence of one of these bosons from the non-observation of the other, or conversely, if given the observation of one of them we could get guidance in the search for the other one. In other words, we aim to unravel the phenomenological connection between the two extra bosons. We provide information on our computational setup in Section~\ref{CompSetup}, and study correlations among cross sections and asymmetries in Sections~\ref{sec:xsections} and \ref{sec:asym} respectively.

%%%%%%%%%%%%%%%%%%%%%%%%%
\subsection{Computational setup} \label{CompSetup}
To perform our numerical analysis, we have implemented the models presented in Section~\ref{sec:theo} into {\sc FeynRules} (version 2.3.35)~\cite{Alloul:2013bka} and converted them into leading-order UFO libraries~\cite{Christensen:2009jx,Degrande:2011ua}. Our new implementations are based on already publicly available ones~\cite{Mattelaer:2016ynf,Fuks:2017vtl,Borah:2018yxd,Frank:2018ifw,Frank:2019nid}, that we have harmonised with each other in the context of simplified versions of the various models (that are sufficient for our purpose). Hard-scattering parton-level events have then been generated with {\sc MG5aMC} (version 3.0.3)~\cite{Alwall:2014hca}, and have been analysed with {\sc MadAnalysis 5}~\cite{Conte:2012fm,Conte:2014zja,Conte:2018vmg} (version 1.8.58). In our simulations, we convoluted the relevant hard-scattering matrix elements with the leading-order set of NNPDF~3.1~\cite{Ball:2017nwa} parton distribution functions (PDF), that has been accessed through {\sc LHAPDF}~\cite{Buckley:2014ana}.

%%%%%%%%%%%%%%%%%%%%%%%%
\subsection{Production cross sections}\label{sec:xsections}
%%%%%%%%%%%%%%%%%%%%%%%%%

\begin{figure}
 \centering
 \includegraphics[width=0.50\textwidth]{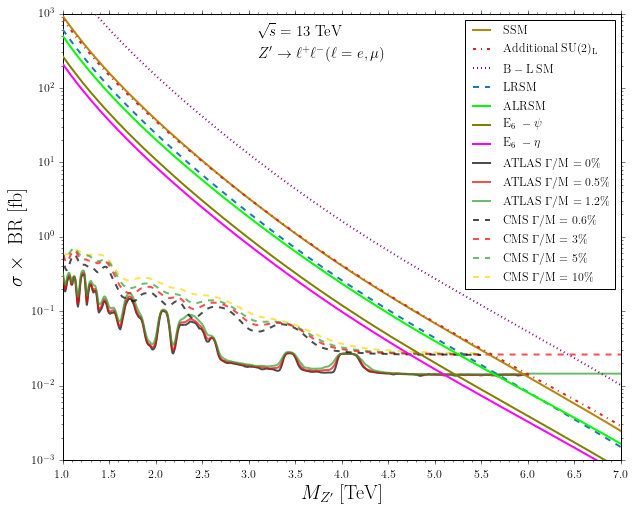}
 \caption{$Z^\prime$ production cross sections, as a function of the $Z'$ boson mass and at  the $\sqrt{s}=13$~TeV LHC. We consider a $Z'$ decay into a dilepton final state, and we show predictions for various scenarios and models. These predictions are compared to current CMS and ATLAS limits. In this figure, we ignore the uncertainties on the theory results for clarity, as those are discussed in Figure~\ref{fig:ZpLimits}.} \label{fig:LimitsInGeneral}
\end{figure}

We begin by investigating, in Figure~\ref{fig:LimitsInGeneral}, the existing and relevant $Z^\prime$ limits originating from LHC searches at a centre-of-mass energy of $\sqrt{s}=13$~TeV. We compute cross section predictions associated with the production of a dileptonically-decaying $Z'$ boson in each model considered. Our predictions are then compared with the most recent ATLAS~\cite{ATLAS:2019erb} and CMS~\cite{Sirunyan:2021khd} bounds, that we display for various assumptions on the width-over-mass ratio of the $Z'$ boson, $\Gamma_{Z'}/M_{Z'}$. All adopted values of this ratio are broadly compatible with those emerging from the BSM scenarios under consideration. In the figure, we abusively make use of the $\sigma\times {\rm BR}$ notation for the title of the $y$-axis for simplicity, although our calculations are related to the full process $p p \to \gamma, Z,Z'\to \ell^+\ell^-$. The cross section therefore includes both the SM contribution (squared), the new physics contribution (squared), and their interference. The pure SM component of the cross section (representing the SM background) is generally small, especially for the dilepton invariant mass ($M_{\ell\ell}$) intervals typically considered in the corresponding experimental analyses. Moreover, the new physics states are generically quite narrow (see Table~\ref{tab:WZprimeparam}  for examples), so that a factorised approach in the NWA would give production rates times branching ratio values compatible with those arising from the complete calculation. Whilst this is true for integrated cross sections, this may not hold at the differential cross section level, as distributions remain subject to non-negligible interference effects enabled by a finite value of the new gauge boson width.

The lowest $Z^\prime$ mass limits are found for the $E_6$-inspired $\psi$ and $\eta$ models (defined in Table~\ref{tab:u1charges}), in which we get $M_{Z'} \gtrsim 5$~TeV. In a left-right-symmetric setup ({\it i.e.}\ for the ALRSM and LRSM models), we have found slightly stronger bounds with $M_{Z'} \gtrsim 5.6$~TeV. Finally, models including an additional $SU(2)_L$ symmetry or a $B-L$ extra gauge symmetry feature larger cross sections, which yield even stronger limits with $M_{Z'} \gtrsim 6$~TeV and 6.7~TeV respectively. The hierarchy between those different bounds show that there is no generic impact of the presence of an additional charged gauge boson in the model. The bounds are indeed model-dependent and cannot be generically extracted. This justifies our model-by-model approach.

\begin{figure}
	\centering
	\includegraphics[width=0.48\textwidth]{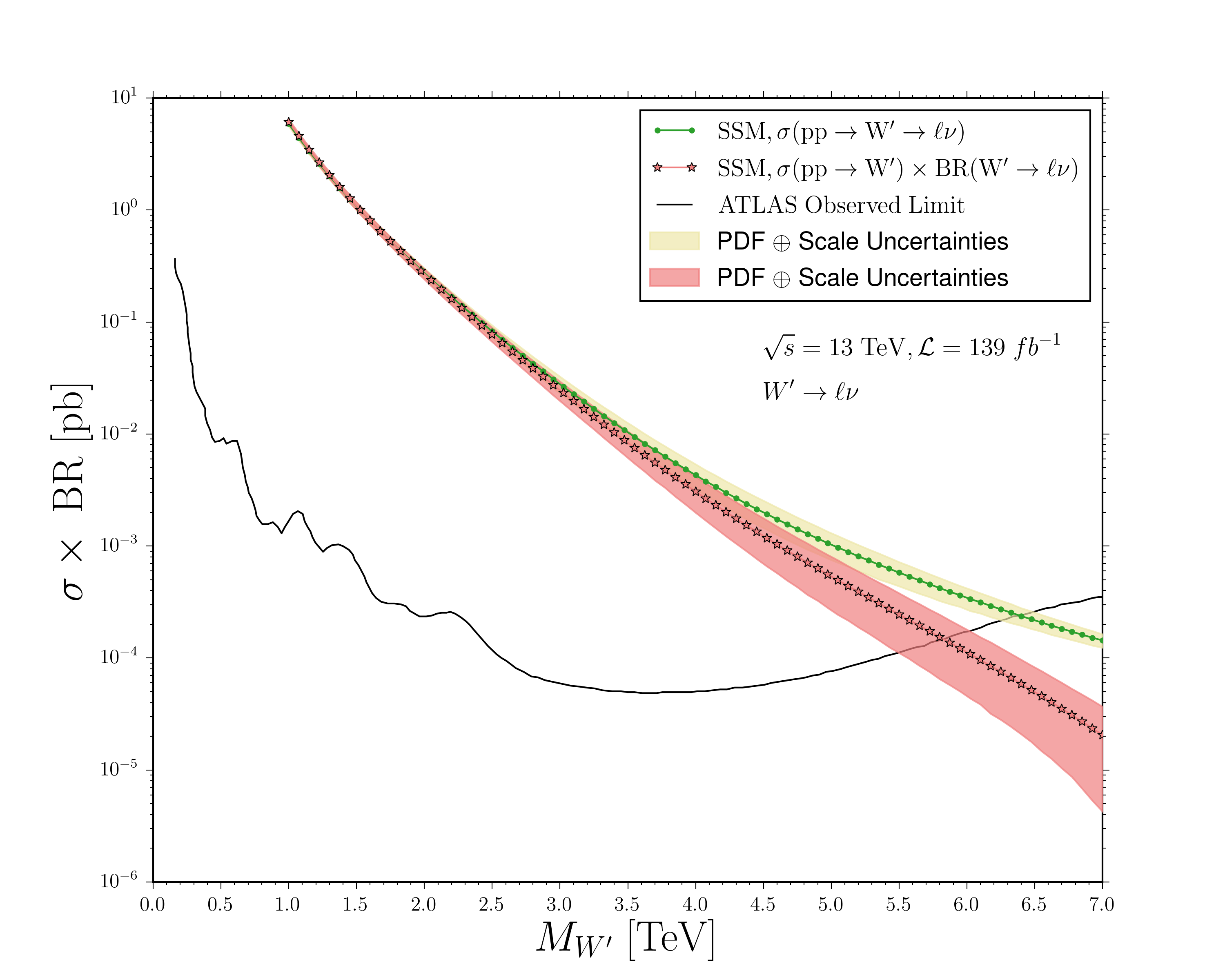}
	\includegraphics[width=0.48\textwidth]{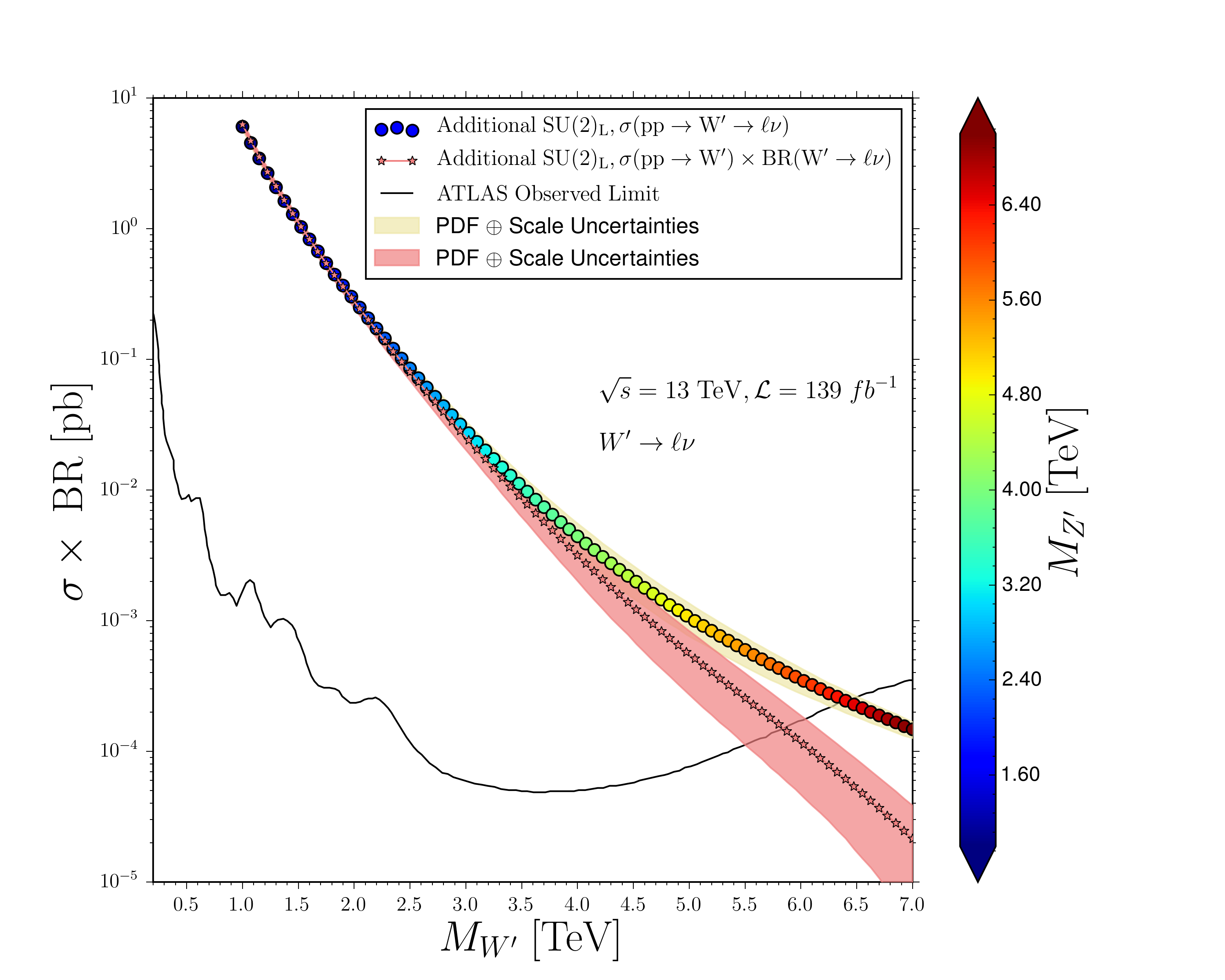} 		   
	\caption{$W^\prime$ production cross section as a function of the $W'$ boson mass and at the $\sqrt{s}=13$~TeV LHC. We consider a $W'$ boson decaying into a charged lepton plus neutrino final state, and we show predictions for various scenarios in the SSM (left) and for models with an additional $SU(2)_L$ symmetry (right).  The shaded areas include a combination in quadrature of the PDF and scale uncertainties, and for models with an additional $SU(2)_L$ symmetry, the colour code indicates, when it is used, the mass of the accompanying $Z^\prime$ state for each considered $M_{W'}$ value. In both figures, the predictions are compared to the current experimental bounds. The two pairs of curves compare the bounds obtained when a full calculation of the matrix element related to the $pp\to W' \to \ell\nu$ is used, and when the NWA is used (so that the production cross section factorises from the decay matrix element).}
	\label{fig:LimitsInGeneral2}
\end{figure}

We repeat the exercise in Figure~\ref{fig:LimitsInGeneral2} for $W'$ boson production at the LHC, when the extra gauge boson decays into an associated charged-lepton/neutrino pair. We consider SSM predictions (left panel), as well as predictions in models featuring an additional $SU(2)_L$ symmetry (right panel), and the predictions are compared in both cases to current ATLAS bounds~\cite{ATLAS:2019lsy}.  We do not show any result for the LRSM and ALRSM scenarios, as we cannot extract limits on the $W'$ properties from the CC Drell-Yan channel in those models. The $W^\prime$ boson indeed always decays into a final state different from the $\ell\nu$ one.

In the setup featuring an extra $SU(2)_L$ symmetry, the $W'$ boson is always accompanied by a $Z'$ boson whose mass, that is not an independent parameter (see Eq.~(\ref{eq:ZpWpmassLL})), is represented through a colour code for each $M_{W'}$ value. This is one of the relevant points emphasised in this study: we advocate that bounds on one gauge boson can imply potentially stronger bounds (like here) on the other one (for models featuring both new gauge bosons). 
Here, direct searches for an $SU(2)_L$ $Z'$ gauge boson set limits of 6~TeV on its mass, while the associated $W'$ searches imply more stringent bounds of $M_{Z'} \gtrsim 6.4$~TeV. Likewise, if with an increased luminosity an excess were to be found in either of the two search channels, the search in the other channel would be in the ideal position to optimise the mass selection to evidence the presence of the companion new gauge state. Clearly, the same approach cannot be exploited in the case of the SSM in which the two masses are independent parameters by construction.

Figure~\ref{fig:LimitsInGeneral2} also compares the bounds that could be derived when relying on a computation of the full matrix element associated with the $pp\to W'\to \ell \nu$ process with those obtained when this matrix element is approximated by its NWA form. In this last case, off-shell effects are lost and the cross section gets reduced by a factor of a few in the large-mass regime. Consequently, the bounds on the extra gauge boson are weaker. Moreover, allowing the $W'$ boson to be offshell implies that a phase-space region associated with a smaller Bjorken-$x$ could contribute to the cross section, even dominantly like here due to larger PDF values. This additionally yields, in this case, smaller PDF uncertainties. Handling a full calculation offers hence a possibility to extract stronger and more robust bounds through a greater cross section and a better control of the theory errors. For that reason, all the results below are extracted from predictions for the full process.

\begin{figure}
	\centering
	\includegraphics[width=0.48\textwidth]{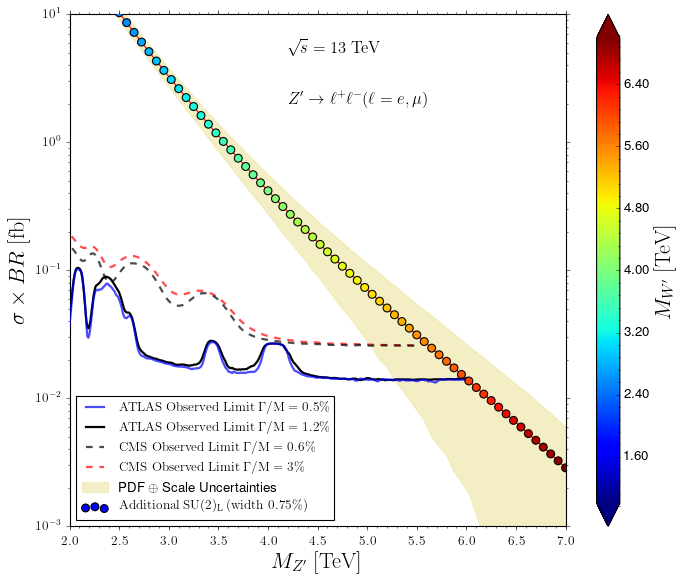} \\
        \includegraphics[width=0.48\textwidth]{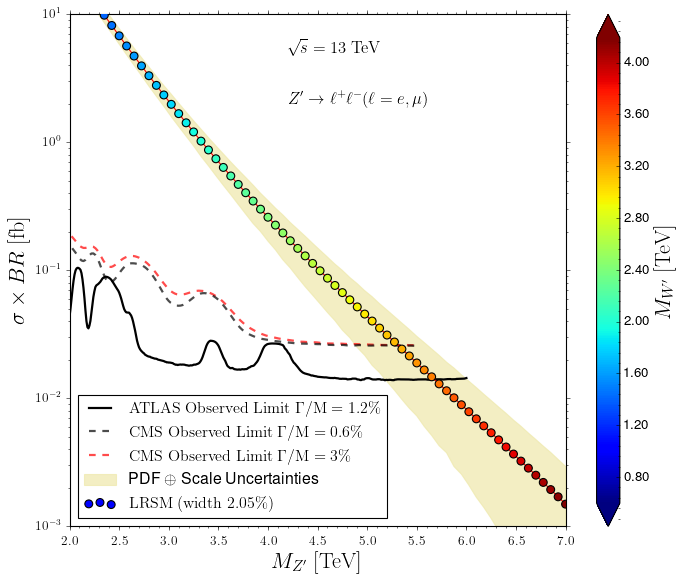}\hfill
	\includegraphics[width=0.48\textwidth]{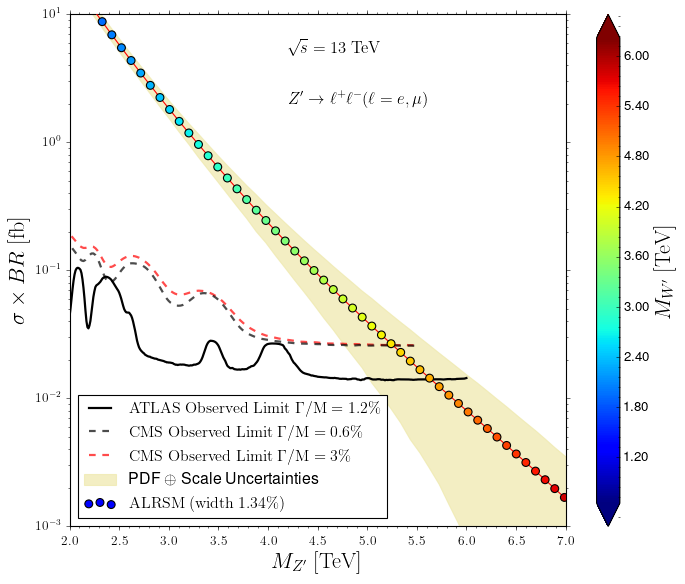}
	\caption{Same as in Figure~\ref{fig:LimitsInGeneral} but for individual models considered separately. We display results for BSM setups with an additional $SU(2)_L$ symmetry (top row), and for the LRSM (bottom left) and ALRSM (bottom right). Our predictions are provided together with the associated PDF and scale uncertainties given as shaded areas, and the colour code associated with each $Z'$ mass choice refers to the mass of the accompanying $W'$ state.}
	\label{fig:ZpLimits}
\end{figure}

In Figure~\ref{fig:ZpLimits}, we revisit the $Z'$ bounds investigated in Figure~\ref{fig:LimitsInGeneral}, this time including theory errors on the predictions and considering all models that feature both a $W'$ and a $Z'$ boson. We display information about the experimental limits on the $Z'$ boson, and about the mass of the associated $W'$ boson that we provide through a colour code. As before, the obtained $Z'$ bounds allow for the extraction of a limit on the $W'$ mass. For example, $Z'$ mass limits in the ALRSM and the LRSM scenarios imply that the corresponding $W'$ boson mass has to be of at least 4.8~TeV and 3.5~TeV respectively. This therefore provides strong $W'$ bounds despite the absence of any direct search limits in the ALRSM and LRSM.

The constraining power in the high mass regime is found to be poorer than expected because of the theoretical errors, and in particular because of the PDF errors that grow quite significantly with $M_{Z'}$. Such an issue was also present in the CC case (as illustrated in Figure~\ref{fig:LimitsInGeneral2}) as well as for $U(1)'$ models for which predictions for $Z'$ production are similar to those shown in Figure~\ref{fig:ZpLimits}. The impact of this growth of the errors may however be tamed down by the different masses of the two extra bosons. An assessment of the likelihood for the heavier boson to exist in LHC data can indeed accurately be ascertain through a search for the lighter boson, as in this case the parton density errors inherent to the calculation are smaller. These larger PDF errors in the large Bjorken-$x$ regime are connected to the way the NNPDF sets are extracted, and the usage of CT \cite{Hou_2021} or MSHT  \cite{Bailey_2021} densities
would lead to a more reasonable high-mass behaviour. This however comes at the cost of a theoretical bias on the functional form of the individual PDF, as already pointed out in the context of the Minimal Supersymmetric Standard Model and for leptoquark scenarios~\cite{Frixione:2019fxg,Borschensky:2020hot,Borschensky:2021hbo}. As it is conceivable that the modelling of the structure functions will see improvements in the years to come, it is quite likely that both channels will benefit from them in a similar way. The advocated cross-fertilising approach will thus be even more in order and valid.

Given the current constraints on the $W'$ and $Z'$ masses, it is rather unlikely that the upcoming Run 3 of the LHC can yield significant advances relative to where we stand today. We therefore decide to make use of the expected increased luminosity of the future high-luminosity phase of the LHC ({\it i.e.}\ the HL-LHC \cite{Gianotti:2002xx,CidVidal:2018eel}) to establish whether cross-correlations between the $W'$ and $Z'$ bounds may exist in the case of various asymmetries, whose extraction requires more substantial event rates. We benchmark our forthcoming results within the scope of the LHC operating at 14~TeV and with a total luminosity of 3000~fb$^{-1}$, and we rely on this tenfold increase in event rates with respect to those achievable during the LHC Run 3 to study asymmetries in the considered Drell-Yan channels. We examine those asymmetries in the next subsection, and study them as a function of the invariant mass $M_{\ell\ell}$ of the lepton pair in the NC case, and as a function of the transverse mass $M_T$ of the $\ell\nu$ system in the CC  case.

\subsection{Asymmetries}\label{sec:asym}

\begin{table}\centering
  \renewcommand{\arraystretch}{1.6}
  \setlength\tabcolsep{6pt}
  \begin{tabular}{c| cccc | cc}
    Models& SSM & LRSM & ALRSM  & $SU(2)_L'$ & $U(1)_\psi$  & $U(1)_\eta$  \\ \hline
    $M_{Z^\prime}$ & 6~TeV & 6~TeV & 6~TeV  & 6~TeV & 5.5~TeV  & 5.5~TeV \\
    $\Gamma_{Z^\prime} / M_{Z^\prime}$& 2.95\% & 2.05\% & 1.34\% & 0.75\% & 0.59\%  & 1.02\%\\ \hline
    $M_{W^\prime}$ & 6.5~TeV &6~TeV & 5~TeV & 6.5~TeV & - & -  \\
    $\Gamma_{W^\prime} / M_{W^\prime}$& 0.30\%  & 0.39\% & 0.34\% & 0.30\% & - & - 
  \end{tabular}
	\caption{Extra gauge boson masses and widths adopted in our study.}
\label{tab:WZprimeparam}\end{table}

In this section, we study various asymmetries arising from extra gauge boson production through Drell-Yan processes at the LHC. We fix the new gauge boson masses in the different scenarios as shown in Table~\ref{tab:WZprimeparam}, these masses being chosen as the lowest allowed ones in a given model according to the constraints depicted in Sections~\ref{sec:WpZpmass} and \ref{sec:xsections}. The widths $\Gamma_{Z'}$ and $\Gamma_{W'}$ are then evaluated analytically.

We begin by focusing on the NC Drell-Yan channel for which we consider the forward-backward asymmetry $A_{\rm FB}^{\rm True}$ that we define by
\begin{equation}
  A^{\rm True}_{\rm FB}= \frac
   {\frac{{\rm d} \sigma}{{\rm d}M_{\ell\ell}}(\cos\theta_{\ell^-}>0)-\frac{{\rm d} \sigma}{{\rm d}M_{\ell\ell}}(\cos\theta_{\ell^-}<0)}
   {\frac{{\rm d} \sigma}{{\rm d}M_{\ell\ell}}(\cos\theta_{\ell^-}>0)+\frac{{\rm d} \sigma}{{\rm d}M_{\ell\ell}}(\cos\theta_{\ell^-}<0)}\, ,
\end{equation}
where $\theta_{\ell^-}$ is the negatively-charged lepton polar angle relative to the initial-quark direction axis. While such an angle allows for the identification of the forward and backward hemispheres, the quark direction cannot be obtained without relying on Monte Carlo information. For this reason, we additionally consider the reconstructed forward-backward asymmetry $A_{\rm FB}^{\rm Rec}$ (sometimes noted as $A{^*_{\rm FB}}$) that is defined from the rapidity of the dilepton system $y_{\ell\ell}$,
\begin{equation}	
 y_{\ell{\ell}} = \frac{1}{2} \ln \left[  \frac{E_{\ell\ell} + p^z_{\ell\ell}}{E_{\ell\ell} - p^z_{\ell\ell}} \right].
\end{equation}
In this expression, $E_{\ell\ell}$ and $ p^z_{\ell\ell}$ stand for the energy and the momentum in the $z$-direction of the lepton pair respectively. The $y_{\ell\ell}$ variable provides a measure of the boost induced by the initial state partonic configuration, and its sign allows for a definition of the direction of the initial quark. The polar angle is then extracted in the dilepton restframe. In the following, we examine the validity of the approximation $A_{\rm FB}^{\rm Rec} \approx A^{\rm True}_{\rm FB}$, that gets better once an additional cut on $|y_{\ell\ell}|$ is implemented. We consider several options for the corresponding lower thresholds, that are varied from 0 (no cut) to 0.6.

\begin{figure}
  \centering
  \includegraphics[width=0.495\textwidth,trim={1cm 1cm 2cm 1.5cm},clip]{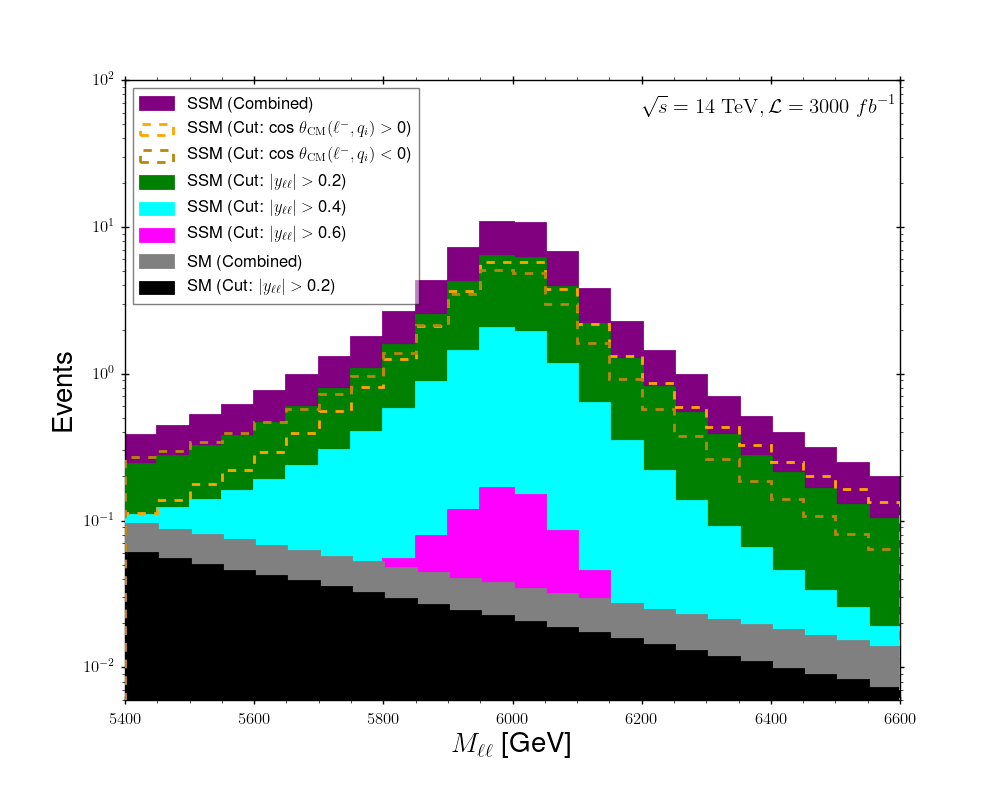}\hfill
  \includegraphics[width=0.490\textwidth]{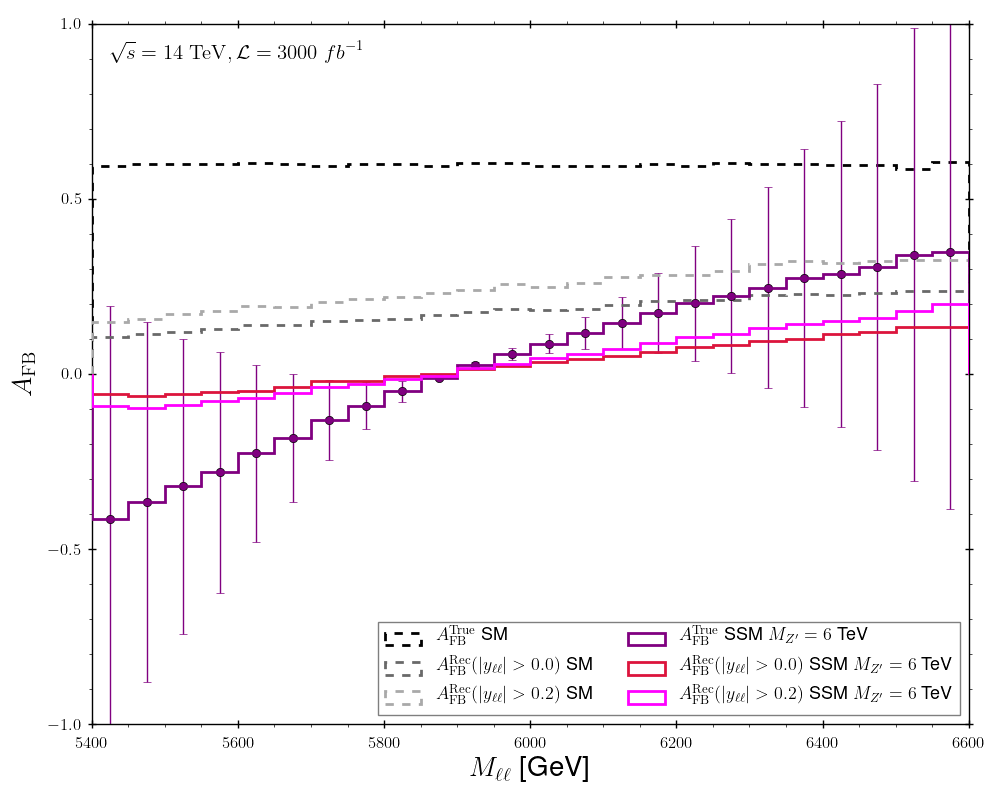}
  \caption{Distribution in the invariant mass $M_{\ell\ell}$ of the lepton pair produced in SM and SSM NC Drell-Yan processes at the HL-LHC (left panel). The dashed lines represent events with $\cos\theta_{\ell^-} > 0$ and $\cos\theta_{\ell^-} < 0$, and the solid areas are associated with the spectra obtained after the implementation of various $|y_{\ell\ell}|$ cuts. Corresponding true and reconstructed forward-backward asymmetries are shown in the right panel of the figure. We refer to the text for more details.}
  \label{fig:AFB_SSM}
\end{figure}

\begin{figure}
	\centering
	\includegraphics[width=0.485\textwidth]{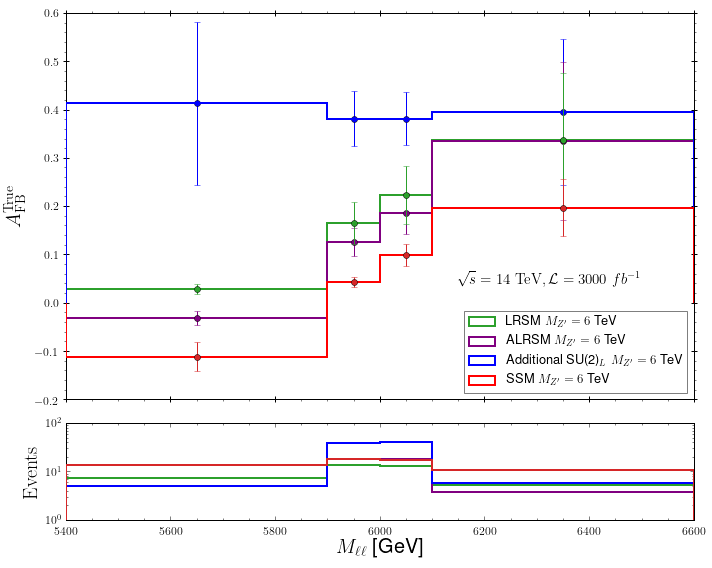}\hfill
	\includegraphics[width=0.485\textwidth]{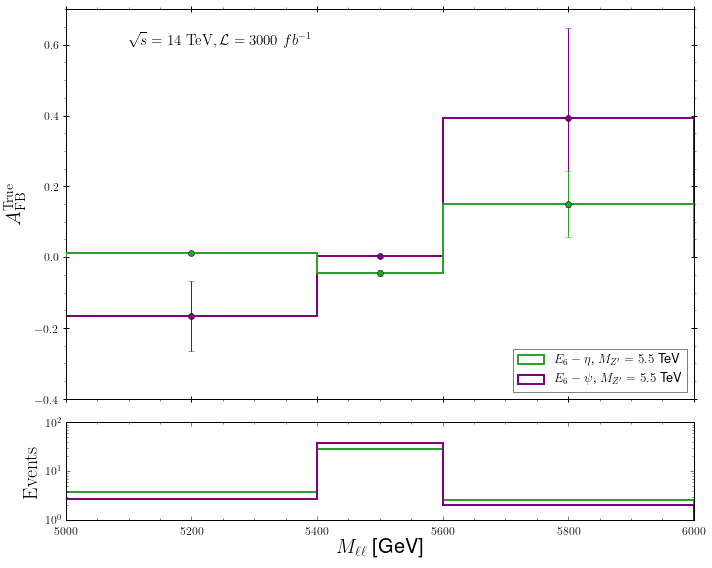}\\ %trim={0.70cm 0 2cm 0}, clip
	\includegraphics[width=0.485\textwidth]{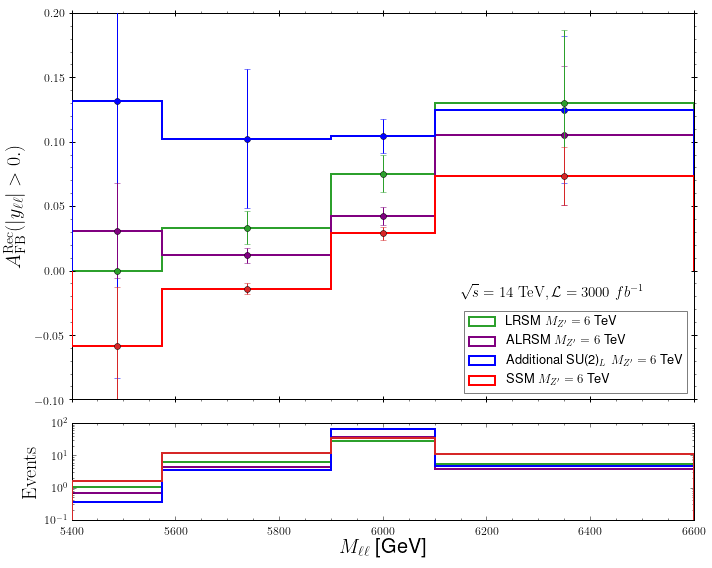}\hfill
	\includegraphics[width=0.485\textwidth]{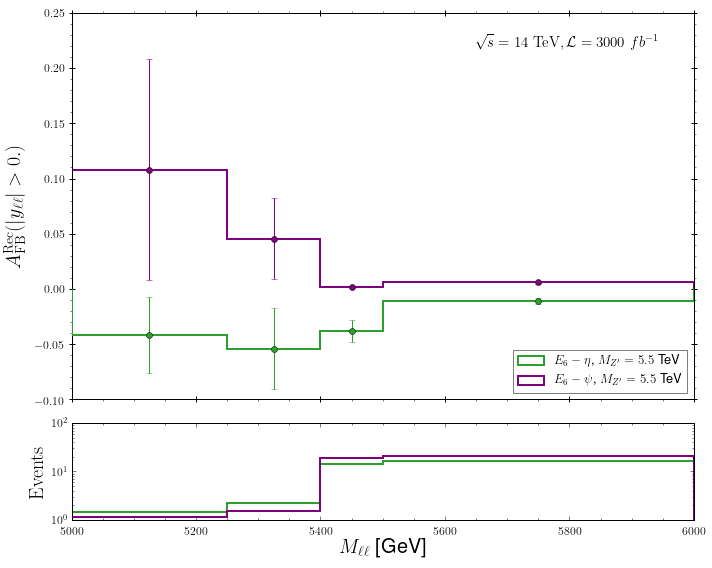}
	\caption{Dependence of the forward-backward asymmetry on the invariant mass $M_{\ell\ell}$ of the lepton pair produced in NC Drell-Yan processes at the HL-LHC, in models featuring both a $W'$ and a $Z'$ boson (left panel), and in $U(1)'$ models (right panel). Predictions are shown for the true (top row) and reconstructed (bottom row) asymmetries.  In the lower panels of each subfigure, we show the number of events populating a given bin, which is related to the associated uncertainty according to Eq.~\eqref{eq:errAFB}. 
 \label{fig:AFB}}
\end{figure}

For every bin in $M_{\ell\ell}$ we compute the associated statistical uncertainty $\delta A_{\rm FB}$ on the asymmetry,
\begin{equation}\label{eq:errAFB}
	\delta A_{\rm FB} = \sqrt{ \frac{4}{ \mathcal{L}} \frac{\sigma_F \sigma_B}{ (\sigma_F + \sigma_B)^3}} = \sqrt{\frac{1 - A_{\rm FB}^2}{ \sigma \mathcal{L}}} =  \sqrt{\frac{1 - A_{\rm FB}^2}{ N }},
 \end{equation}
where $\mathcal{L}$ is the integrated luminosity, $N$ is the total number of events populating the considered $M_{\ell\ell}$ bin (associated with the cross section $\sigma$), and $A_{\rm FB}$ equivalently stands for the true or reconstructed forward-backward asymmetry. Moreover, we denote by $\sigma_F$ and $\sigma_B$ the forward and backward cross sections in the bin. Our results are given in Figure~\ref{fig:AFB_SSM} for the SSM, and in Figure~\ref{fig:AFB} for all models featuring both a $W'$ and a $Z'$ boson (left panel), and for $U(1)'$ models (right panel). Figure~\ref{fig:AFB_SSM} additionally includes the corresponding SM expectation as a reference.

We provide, in the left panel of Figure~\ref{fig:AFB_SSM}, predictions for the $M_{\ell\ell}$ distributions in the SM and the SSM after several cuts on the dilepton rapidity $|y_{\ell\ell}|$ (as filled histograms). Our results show that a too strict cut on the $|y_{\ell\ell}|$ variable is statistically not an option. Too few events would indeed survive the cut, implying that there is not enough statistics to meaningfully compute $A_{\rm FB}$ asymmetries, and therefore verify that $A_{\rm FB}^{\rm True}\approx A_{\rm FB}^{\rm Rec}$ after the cut. Such a cut, although potentially useful, is therefore ignored in the following. We additionally present, for the SSM, the resulting invariant mass distributions when we solely consider events populating the forward ($\cos\theta_{\ell^-}>0$; orange dashes) and backward ($\cos\theta_{\ell^-}<0$; brown dashes) hemispheres. This clearly demonstrates the existence of an asymmetry, that is studied in the right panel of the figure.

The asymmetry results are generally plagued with very large uncertainties, that we only show for the SSM case without any $|y_{\ell\ell}|$ cut for clarity. This consists in the most optimistic situation from the point of view of the errors. The SM background is indeed by definition extremely low, as one lies in a very large $M_{\ell\ell}$ regime, and the SSM signal is quite reduced once cuts are implemented. By taking errors into account, we can observe that the $A_{\rm FB}^{\rm True}$ and $A_{\rm FB}^{\rm Rec}$ distributions are almost compatible with each other, at least when no $|y_{\ell\ell}|$ cut is included. The issue with the size of the errors can be addressed by focusing only on the $Z'$ peak where statistics is more significant. This allows for a precise (enough) determination of the shape of the asymmetry distribution, so that it could be used for signal characterisation and to distinguish various $Z'$ signals at a given mass (see below). While we could in principle also compare our new physics signal to the SM predictions, we refrain from doing so. The SM asymmetry is indeed plagued with enormous statistical errors (not shown on the figure), preventing us from making any meaningful conclusive statement. This is neverthless not crucial as the invariant mass spectrum (left panel of Figure~\ref{fig:AFB_SSM}) is sufficient to observe a sign of an extra gauge boson relative to the SM. 

However, observing a $Z'$ state at a given mass (of 6~TeV here) does not guarantee that we can get handles on the underlying theory. In the following, we try to examine how using asymmetries in the $M_{\ell\ell}$ region close to the peak (where we have in principle enough statistics) could help.
In Figure~\ref{fig:AFB}, we show the forward-backward asymmetries resulting from Drell-Yan dilepton production in models featuring additional $W'$ and $Z'$ bosons (left panel), and in $U(1)'$ models (right panel). We provide our BSM predictions together with the associated statistical uncertainties. The overarching message emerging from the results is that alongside the total rate, the $A_{\rm FB}$ observable is useful for the separation of one model from another. While this is clear for most pairs of models when relying on the true $A_{\rm FB}^{\rm True}$ spectra, the discriminating power remains intact for several pairs of models when using reconstructed $A_{\rm FB}^{\rm Rec}$ spectra. However, due to the limited number of events, some care should be given to optimising the binning. Statistical uncertainties are indeed rather large away from the $Z'$ mass position, while they can well be controllable near it. In the light of this, it will be important to also combine ATLAS and CMS data samples to increase the statistics and reduce the uncertainties.

\begin{figure}
	\centering
	\includegraphics[width=0.51\textwidth]{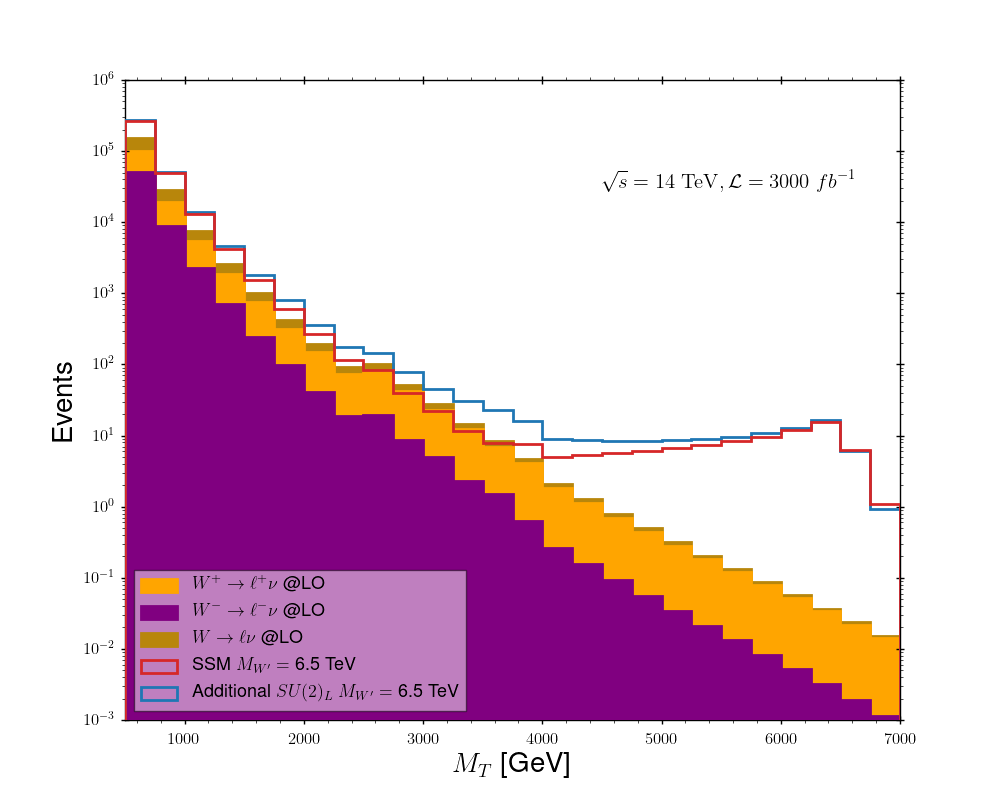}\hfill
	\includegraphics[width=0.485\textwidth]{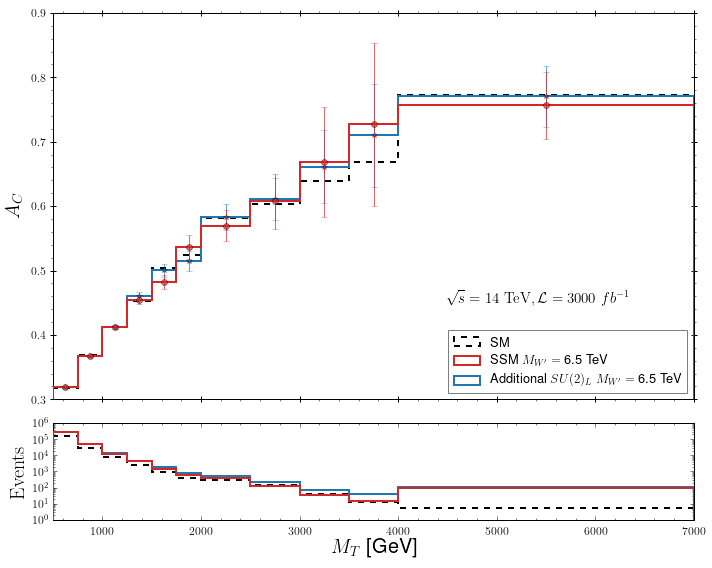}
  \caption{Distribution in the transverse mass $M_T$ of the $\ell\nu$ system produced in CC Drell-Yan processes at the HL-LHC (left panel). We show predictions for the SM (olive), as well as for the SSM (red) and for models with an additional $SU(2)_L$ symmetry (blue). In the case of the SM, we distinguish the production of a positively-charged (purple) and negatively-charged (yellow) $\ell\nu$ system. The corresponding charge asymmetries are shown in the right panel of the figure, together with the associated statistical uncertainties.  In the lower panel of the right subfigure,  we show the number of events populating a given bin, which is related to the associated uncertainty according to Eq.~\eqref{eq:errAFB}. 
  \label{fig:AC}}
\end{figure}

To conclude, we have shown that in the case of an excess in the NC Drell-Yan channel, its nature could be diagnosed in terms of the chiral couplings of the underlying $Z'$ state through the usage of asymmetries. This could then in turn be used (through the existence of $W'$ and $Z'$ correlations in models containing both bosons) to predict the shape of the asymmetry that could emerge in the CC Drell-Yan mode $pp\to \ell\nu$, to which we turn next. We consider in this case the lepton charge asymmetry $A_{\rm C}$,
\begin{equation}
  A_{\rm C}=\frac
  {\frac{{\rm d} \sigma}{{\rm d}\eta_\ell}(pp\to\ell^+\nu) - \frac{{\rm d} \sigma}{{\rm d}\eta_\ell}(pp\to\ell^-\bar\nu)}
  {\frac{{\rm d} \sigma}{{\rm d}\eta_\ell}(pp\to\ell^+\nu) + \frac{{\rm d} \sigma}{{\rm d}\eta_\ell}(pp\to\ell^-\bar\nu)}\, ,
\end{equation}
in which the variable $\eta_\ell$ refers to the final-state charged lepton pseudorapidity. We additionally include statistical uncertainties in the calculation of the charge asymmetry, that we derive similarly to Eq.~\eqref{eq:errAFB}.

The results are presented for the relevant models ({\it i.e.}\ for the SSM and $SU(2)_L'$ models) in Figure~\ref{fig:AC}, in which we focus on the distribution in the transverse mass $M_T$ of the $\ell\nu$ system (left panel) and on the corresponding charge asymmetries (right panel). In the left panel, we present $M_T$ spectra for the SM (olive histogram), as well as for the SSM (red) and the $SU(2)_L'$ case (blue). As already found in the context of the $M_{\ell\ell}$ distribution for neutral currents, the $M_T$ distribution arising in charged currents is sufficient to observe deviations from the SM through the presence of a significant peak around $M_W'$, such a peak being absent in the SM. By virtue of the parton densities, a charge asymmetry exist, as shown in the SM through the purple (the $pp\to \ell^-\bar\nu$ process) and yellow (the $pp\to \ell^+\nu$ process) histograms. The difference between these two distributions, that is similar in the BSM configurations considered and therefore not shown for the SSM and $SU(2)_L'$ predictions, results from the different parton densities relevant for the two hard-scattering processes. Consequently, we expect the lepton charge asymmetry to be quite similar in all models, as testified by the right panel of the figure. While the event rate is typically larger than in the NC case (so that the resulting statistical uncertainties are smaller), the discriminative power is typically (much) lower. Therefore, separating the SSM hypothesis from the one assuming a model with an additional $SU(2)_L$ would require to rely essentially on NC events.

%%%%%%%%%%%%%%%%%%%%%%%%%%%%%%%%%%%%%%%%%%%%%%%%%%%%%%
\section{Summary and conclusions}
\label{sec:conclusion}
In this work, we have tried to establish a connection between analyses searching for $W^\prime$ and $Z^\prime$ boson production and decay at the LHC. While only neutral gauge bosons are required by models with additional $U(1)$ groups, both charged and neutral vector states are present in theories with additional $SU(2)$ factors. Correlations may thus exist. Of course, searches for such new gauge bosons have been undertaken already at the LHC, and stringent constraints have been imposed on their properties. The results of these searches can be translated into mass bounds on the extra bosons currently falling in the 5--6~TeV region for typically weak gauge couplings. What is currently missing, though, are correlated searches for charged and neutral gauge bosons. Within a specific model, relevant information about one type of boson can indeed yield information about the other, possibly avoiding the necessity for separate searches or opening the door to more sensitive combined analyses.

Unlike previous literature, we have chosen specific (rather than general or simplified) model implementations to illustrate this point. The advantage is that we can keep the number of parameters to a minimum while taking advantage of the correlations between masses and/or couplings of the two new gauge bosons. Among models with extra charged and neutral gauge bosons, we have selected those with an additional $SU(2)_L$ group, as well as those with an additional $SU(2)_R$ group ({\it i.e.}\ the LRSM and ALRSM). We have moreover compared the associated predictions with those obtained in the SM and the SSM. In our calculations, we have included interference effects with the SM background for all new resonances, taking into account that such high $W'$ and $Z'$ masses can yield a width comparable to the experimental resolution in the invariant and transverse masses of the NC and CC Drell-Yan final states respectively. Furthermore, we have compared our findings with models that include only a single $Z^\prime$ gauge boson, concentrating on those generated by the breaking of the  $E_6$ group, as well as on $U(1)_{B-L}$.

After briefly describing the models considered, we have studied current LHC constraints on the related new charged and neutral gauge bosons. We have then proceeded to study the correlations that could arise from present $W^\prime$ and $Z^\prime$ analyses. In this connection, we have shown how limits on $W'$ signals can be interpreted as constraints on the existence of a $Z'$ boson which are more stringent than those obtained from direct searches. We have further moved on to analyse their Drell-Yan production and decay at the 14~TeV LHC, focusing on a large integrated luminosity of 3000 fb$^{-1}$ like the one attainable at the HL-LHC. Whereas the distributions in both the dilepton invariant mass (for NCs) and the $\ell\nu$ transverse mass (for CCs) differ amongst the considered BSM scenarios, they present themselves in the form of a rather universal Breit-Wigner shape. Getting information on the underlying theory therefore requires a deeper study.

We have proposed to combine the analysis of the differential rates of the cross section with that of the corresponding forward-backward (in the NC case) and lepton charge (in the CC case) asymmetries. We have shown that the asymmetry lineshapes emerging in the different BSM scenarios considered are notably different, in particular for the  $Z'$ analysis. We have therefore concluded that, for both discovery and characterisation purposes, information gathered from $A_{\rm FB}$ studies could be beneficial, providing handles on the underlying model. In our work, we have employed a variable bin size and advocated the use of combined future ATLAS and CMS data samples at the HL-LHC in order to minimise the statistical uncertainties. The limited event rates stemming from the fact that accessible $W'$ and $Z'$ masses are rather heavy indeed hinders somewhat the scope of our procedure at present.

Cross-correlating cross sections and asymmetries appears to be a promising way to distinguish among the possible BSM models embedding both a $W'$ and a $Z'$ boson, and thus mirror the underlying BSM gauge structure. We regard our results as a useful proof-of-concept of a synergetic approach, which validation will require dedicated experimental analyses across different ATLAS and CMS working groups currently engaged in parallel, but  not connected, searches for neutral and charged exotic particles. In this respect, while our choice of new gauge boson masses near their experimentally allowed minimum values at present was done to maximise the sensitivity at the HL-LHC, we are confident that, even after the LHC Run 3, the scope of the advocated analyses will not be significantly reduced. We then consider these as an additional motivation for the deployment of the HL-LHC, though the limited event rate remains a concern which will need to be compensated by ever improving the phenomenological analyses, combining the advances coming from more precise theoretical predictions with those stemming from enhanced experimental efficiencies.

%%%%%%%%%%%%%%%%%%%%%%%%%%%%%%%%%%%%%%%%%%%%%%%%%%%%%
\section*{Code availability}
	%%%%%%%%%%%%%%%%%%%%%%%%%%%%%%%%%%%%%%%%%%%%%%%%%%%%%%

The code in this work can be found in: \url{https://github.com/oozdal/Asymmetries_MA5}

%%%%%%%%%%%%%%%%%%%%%%%%%%%%%%%%%%%%%%%%%%%%%%%%%%%%%
\begin{acknowledgments}
%%%%%%%%%%%%%%%%%%%%%%%%%%%%%%%%%%%%%%%%%%%%%%%%%%%%%%
 M.F. and \"{O}.\"{O}.  thank NSERC for partial financial support under grant number SAP105354. The work of SM is supported in part through the NExT Institute and the STFC Consolidated Grant No.~ST/L000296/1. Parts  of  the  numerical  calculations  reported  in  this  paper  were  performed  using  High  Performance  Computing (HPC) managed by Calcul Quebec and Compute Canada, and the IRIDIS High Performance Computing Facility and associated support services at the University of Southampton.
\end{acknowledgments}
%%%%%%%%%%%%%%%%%%%%%%%%%%%%%%%%%%%%%%%%%%%%%%%%%%%%%%%%%%%%%

\bibliography{WZinterference}

\end{document}